\documentclass[twoside,twocolumn,9pt]{article}
\usepackage{extsizes}
\usepackage[super,sort&compress,comma]{natbib} 
\usepackage[version=3]{mhchem}
\usepackage[left=1.5cm, right=1.5cm, top=1.785cm, bottom=2.0cm]{geometry}
\usepackage{balance}
\usepackage{widetext}
\usepackage{times,mathptmx}
\usepackage{sectsty}
\usepackage{graphicx} 
\usepackage{lastpage}
\usepackage[format=plain,justification=raggedright,singlelinecheck=false,font={stretch=1.125,small,sf},labelfont=bf,labelsep=space]{caption}
\usepackage{float}
\usepackage{fancyhdr}
\usepackage{fnpos}
\usepackage[english]{babel}
\usepackage{array}
\usepackage{droidsans}
\usepackage{charter}
\usepackage[T1]{fontenc}
\usepackage[usenames,dvipsnames]{xcolor}
\usepackage{setspace}
\usepackage[compact]{titlesec}

\usepackage[colorlinks,linkcolor={blue},citecolor={blue},urlcolor={blue}]{hyperref}

\usepackage{bm,amsmath}

\usepackage{latexsym}

\usepackage{verbatim}

\usepackage{color}

\usepackage{subfigure}

\usepackage{gensymb}

\usepackage{float}

\usepackage{paralist}
\usepackage[dvipsnames]{xcolor}
\usepackage[utf8]{inputenc}

\usepackage{epstopdf}

\definecolor{cream}{RGB}{222,217,201}

\begin{document}

\pagestyle{fancy}
\thispagestyle{plain}
\fancypagestyle{plain}{

}

\makeFNbottom
\makeatletter
\renewcommand\LARGE{\@setfontsize\LARGE{15pt}{17}}
\renewcommand\Large{\@setfontsize\Large{12pt}{14}}
\renewcommand\large{\@setfontsize\large{10pt}{12}}
\renewcommand\footnotesize{\@setfontsize\footnotesize{7pt}{10}}
\makeatother

\renewcommand{\thefootnote}{\fnsymbol{footnote}}
\renewcommand\footnoterule{\vspace*{1pt}%
\color{cream}\hrule width 3.5in height 0.4pt \color{black}\vspace*{5pt}} 
\setcounter{secnumdepth}{5}

\makeatletter 
\renewcommand\@biblabel[1]{#1}            
\renewcommand\@makefntext[1]%
{\noindent\makebox[0pt][r]{\@thefnmark\,}#1}
\makeatother 
\renewcommand{\figurename}{\small{Fig.}~}
\sectionfont{\sffamily\Large}
\subsectionfont{\normalsize}
\subsubsectionfont{\bf}
\setstretch{1.125} 
\setlength{\skip\footins}{0.8cm}
\setlength{\footnotesep}{0.25cm}
\setlength{\jot}{10pt}
\titlespacing*{\section}{0pt}{4pt}{4pt}
\titlespacing*{\subsection}{0pt}{15pt}{1pt}

\fancyfoot{}
\fancyfoot[RO]{\footnotesize{\sffamily{1--\pageref{LastPage} ~\textbar  \hspace{2pt}\thepage}}}
\fancyfoot[LE]{\footnotesize{\sffamily{\thepage~\textbar\hspace{3.45cm} 1--\pageref{LastPage}}}}
\fancyhead{}
\renewcommand{\headrulewidth}{0pt} 
\renewcommand{\footrulewidth}{0pt}
\setlength{\arrayrulewidth}{1pt}
\setlength{\columnsep}{6.5mm}
\setlength\bibsep{1pt}

\makeatletter 
\newlength{\figrulesep} 
\setlength{\figrulesep}{0.5\textfloatsep} 

\newcommand{\topfigrule}{\vspace*{-1pt}%
\noindent{\color{cream}\rule[-\figrulesep]{\columnwidth}{1.5pt}} }

\newcommand{\botfigrule}{\vspace*{-2pt}%
\noindent{\color{cream}\rule[\figrulesep]{\columnwidth}{1.5pt}} }

\newcommand{\dblfigrule}{\vspace*{-1pt}%
\noindent{\color{cream}\rule[-\figrulesep]{\textwidth}{1.5pt}} }

\makeatother

\newcommand{\nhat}{\hat{\bf n}}
\newcommand{\that}{\hat{\bf t}}

\newcommand{\Nhat}{\hat{\bf N}}

\newcommand{\zhat}{\hat{\bf z}}

\newcommand{\nb}{\mathbf{n}}

\newcommand{\hb}{\mathbf{h}}

\newcommand{\rb}{{\bf r}}
\newcommand{\tb}{{\bf t}}
\newcommand{\eb}{{\bf e}}
\newcommand{\fb}{{\bf f}}
\newcommand{\Fb}{{\bf F}}
\newcommand{\Ib}{{\bf I}}
\newcommand{\mb}{{\bf m}}
\newcommand{\Tb}{{\bf T}}
\newcommand{\Gb}{{\bf G}}

\newcommand{\qb}{{\bf q}}

\newcommand{\gb}{{\bf g}}

\newcommand{\vb}{{\bf v}}

\newcommand{\zb}{{\bf z}}

\newcommand{\xb}{{\bf x}}

\newcommand{\Jb}{{\bf J}}

\newcommand{\sigtens}{\mbox{\boldmath $\sigma$\unboldmath}}
\newcommand{\Omegab}{\mbox{\boldmath $\Omega$\unboldmath}}
\newcommand{\omegab}{\mbox{\boldmath $\omega$\unboldmath}}

\newcommand{\etatens}{\mbox{\boldmath $\eta$\unboldmath}}
\newcommand{\mutens}{\mbox{\boldmath $\mu$\unboldmath}}

\newcommand{\nablab}{\mbox{\boldmath $\nabla$\unboldmath}} 

\newcommand{\Gammab}{\mbox{\boldmath $\Gamma$\unboldmath}} 

\newcommand{\Btilde}{{\tilde B}}

\newcommand{\Ktilde}{{\tilde K}}

\newcommand{\Ctilde}{{\tilde C}}

\mathchardef\mhyphen="2D
\newcommand{\hyphen}{\mathit{\mhyphen}}

\twocolumn[
  \begin{@twocolumnfalse}
\vspace{3cm}
\sffamily
\begin{tabular}{m{4.5cm} p{13.5cm} }

                                 & \noindent\LARGE{\textbf{Dynamics of a
bacterial flagellum under reverse rotation
}}\\
\vspace{0.3cm} & \vspace{0.3cm} \\

 & \noindent\large{Tapan Chandra Adhyapak$^{\ast}$\textit{$^{a}$} and Holger
Stark\textit{$^{a}$}}\\

                                   & \noindent\normalsize{To initiate tumbling
of an \emph{E.~coli}, one of the helical flagella reverses its sense of
rotation.  It then transforms from its normal form first to the transient
semicoiled state and subsequently to the curly-I state.  The dynamics of
polymorphism is effectively modeled by describing flagellar elasticity through
an extended Kirchhoff free energy.  However, the complete landscape of the free
energy remains undetermined because the ground state energies of the
polymorphic forms are not known.  We investigate how variations in these ground
state energies affect the dynamics of a reversely rotated flagellum of a
swimming bacterium. We find that the flagellum exhibits a number of distinct
dynamical states and comprehensively summarize them in a state diagram.  As a
result, we conclude that tuning the landscape of the extended Kirchhoff free
energy alone cannot generate the intermediate full-length semicoiled state.
However, our model suggests an ad hoc method to realize the sequence of
polymorphic states as observed for a real bacterium.  Since the elastic
properties of bacterial flagella are similar, our findings can easily be
extended to other peritrichous bacteria.} \\

\end{tabular}

 \end{@twocolumnfalse} \vspace{0.6cm}

  ]

\renewcommand*\rmdefault{bch}\normalfont\upshape
\rmfamily
\section*{}
\vspace{-1cm}


\footnotetext{\textit{$^{a}$~Institut f\"{u}r Theoretische Physik, Technische
Universit\"{a}t Berlin, Hardenbergstrasse 36, 10623 Berlin, Germany. E-mail:
tapan.c.adhyapak@tu-berlin.de}}








\section{Introduction}
\label{intro}

Although much research has been performed on the mechanical properties of a
prokaryotic or bacterial flagellum and how these properties are related to the
overall bacterial dynamics \cite{berg_compliance, turner2000, darnton2007,
protofilament1, keiichi2010, goldsteinPRL2000, goldsteinPRL2002, wada2008,
lauga_PRL2011, lauga_AnnRev2016, powers2003, reinhardEPJE2010,
reinhardEPJE2012, reinhardPRL2013, gomperSM2012, jawedPRL2015}, still our
understanding of the complex aspects of bacterial locomotion, such as tumbling
of an \emph{E. coli}, remains incomplete \cite{macnab1977, turner2007,
reinhardPRL2013, larson2015}.  The locomotion of bacteria involves rich and
complex physics \cite{raima, luga2009,coretz, coretz2005, saragosti2011,
rodenborn2013, strong, davod2015, stocker2012, gomper_ScRep2015}.  In addition,
experiments on the collective behavior of bacteria have opened new directions
of physics per se, demonstrating groundbreaking phenomena such as turbulence
and superfluidity in living systems \cite{aranson2011, bacterial_turb1,
bacterial_turb2, active_superfluid}. A comprehensive knowledge of bacterial
locomotion is also necessary for a complete understanding of these novel
phenomena.

One main challenge in dealing with the mechanics of the bacterial flagellum is
still to gain a full theoretical understanding of the polymorphic
transformations shown by the flagellum during tumbling \cite{macnab1977,
turner2000}.  The flagellum can exist in different stable polymorphic forms,
transitions among which are induced mechanically, either by the reverse
rotation of the flagellum \cite{macnab1977, turner2007, reinhardPRL2013}, by
the application of stretching forces \cite{darnton2007, reinhardEPJE2010}, or
by external flows \cite{hotani, goldsteinPRL2002}.  Rotation induced
polymorphic transitions occur during the locomotion of the bacterium and affect
the overall bacterial dynamics \cite{macnab1977}. An \emph{E. coli} flagellum,
for example, usually stays in the normal form; but under reverse rotation, it
transforms first into the semicoiled state, and then fully assumes the curly-I
state \cite{berg_ecoli}.

A number of approaches using the Kirchhoff free energy density and extended
versions of it, were proposed to deal with the flagellar multistability
\cite{goldsteinPRL2000, goldsteinPRL2002, wada2008, reinhardEPJE2010}.  In Ref.
\citenum{reinhardPRL2013} an extended Kirchhoff free energy successfully generates
the rotation-induced polymorphic transformations during the locomotion of the
bacterium \cite{reinhardPRL2013}.  However, the landscape of this free energy
with the fixed positions of the local minima and the harmonic shape in their
neighborhood was not fully explored.  Using a linear increase in the ground
state energies of the local minima, the authors were able to demonstrate an
\emph{E.  coli} flagellum transforming from the normal to the curly-I form.
However, the transient semicoiled form in between was not observed.  In our
study here, we explore the flagellar dynamics in the full free energy landscape
by studying the impact of the ground state energies.  Changes in their values
affect the heights of the transition barriers between the local minima
\cite{goldsteinPRL2000, wada2008,reinhardEPJE2010}, which the flagellum has to
pass to locally transform from one polymorphic state to the other. Thus the
full landscape of the free energy determines the flagellar dynamics during
reverse rotation including possible transient and final steady states.

In this paper we present a systematic study of how variations in the unknown
local ground state energies affect the dynamics of a reversely rotated
flagellum attached to a cell body. We study the nature of transitions and the
dynamic stability of polymorphic forms as a function of the barrier heights and
arrive at a comprehensive state diagram that classifies different scenarios of
flagellar dynamics in the parameter space.  In particular, we show that at the
level of the present formulation of flagellar mechanics the experimentally
observed transient semicoiled state, intermediate in the sequence of
polymorphic states during tumbling, cannot be reproduced. However, we suggest
an alternative way within our model to realize the desired sequence of
transitions.

In the following we first outline in sec. \ref{poly_backgrnd} the observations
on a reversely rotated flagellum and introduce the extended Kirchhoff elastic
free energy.  Section \ref{model} summarizes the equations of motion of the
bacterium and their numerical implementation.  Results are then presented in
sec. \ref{results}, followed by a discussion and conclusions in secs.
\ref{disscn} and \ref{concl}, respectively.

\section{Flagellar conformations during reversal and appropriate modeling}
\label{poly_backgrnd}

The reversal of rotating flagella during locomotion has been most elaborately
studied for \emph{E. coli} \cite{turner2000, turner2007, berg_ecoli}. So, in
our study we take \emph{E. coli} as the prototype for flagellated bacteria.
However, since prokaryotic flagella of peritrichous bacteria have a common
molecular structure, which ultimately determines their elasticity
\cite{asakura1969,calladine1978}, we expect our results to be valid for a wide
range of bacteria.              

An \emph{E. coli} has several flagella, which act as propelling units for the
bacterium \cite{berg_ecoli}. The flagellum is a passive helical filament,
connected at one end to a rotary motor embedded in the  cell wall. Most of the
time the motor rotates the flagellum in the counterclockwise sense (as viewed
along the flagellum from its free end toward the cell body) generating the
thrust force for propelling the bacterium.  At this stage the flagellum assumes
the normal form, a left-handed helix of known radius and pitch.  It forms a
bundle with other flagella and thereby defines a unique direction for the net
propulsive force. Frequently, however, the motor reverses and rotates the
flagellum clockwise.  This forces the bundle to disrupt and the cell body to
tumble.  A sequence of polymorphic transitions in the flagellum is observed
\cite{turner2000, turner2007}. In each case these transitions start at the cell
body and proceed towards the free flagellar end. The whole flagellum thus
changes from the normal state, first to the right-handed semicoiled form, and
then finally to the right-handed curly-I form [see Figs.\
\ref{schematic_model}(b) and (d)].

An understanding of the polymorphic states was developed based on the molecular
structure of the flagellum \cite{calladine1975, calladine1976, calladine1978}.
Following this bottom-up approach, several models describe the dynamics of
polymorphism by coarse-graining over the molecular scale
\cite{mesoscopic,reinhardPB, powersPRL2005}.  However, an alternative approach,
where Kirchhoff's continuum theory of an elastic rod was extended to
incorporate multistability, turns out to be simpler and more effective
\cite{goldsteinPRL2000, goldsteinPRL2002, wada_netz_PRL2007}.  One example of
such an extended Kirchhoff free energy reproduces stretching induced
polymorphism most accurately as demonstrated in Ref.\ \citenum{reinhardEPJE2010}
and is also appropriate for modeling rotation-induced polymorphism in a moving
bacterium \cite{reinhardPRL2013}.  We now describe it in more detail.

\subsection{Flagellar elasticity: extended Kirchhoff elastic free energy}
\label{model_elastic} 

\begin{figure}
\begin{center}
\includegraphics[width=1.\columnwidth]{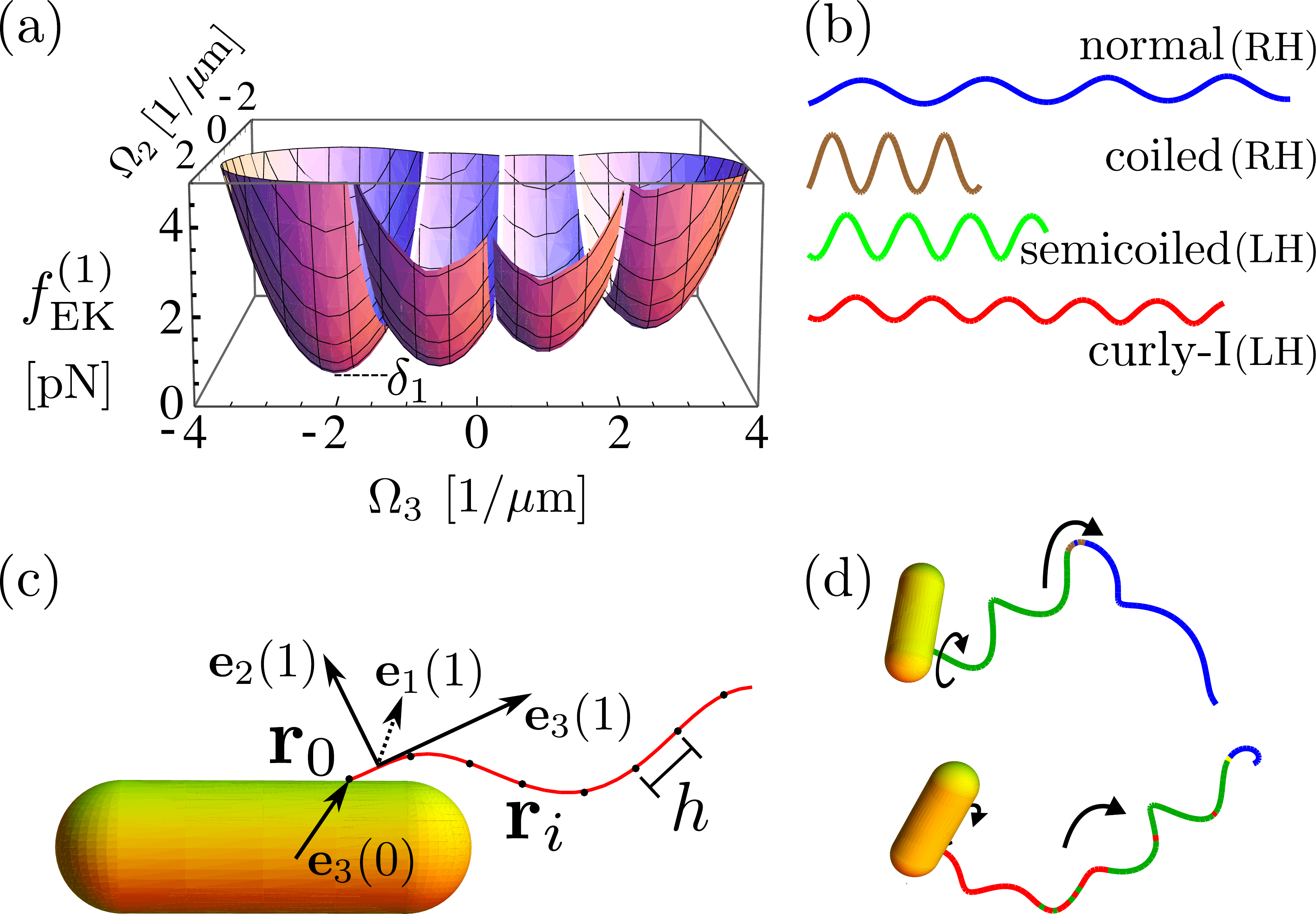}

\caption{(a) The first term of the extended Kirchhoff free energy density,
$f_{\mathrm{EK}}^{(1)}$, plotted against $\Omega_2$ and $\Omega_3$ for
$\Omega_1= 0$.  The local minimum of the normal form with ground state energy
$\delta_1$ is shown. The true local minima for the coiled, semicoiled, and
curly-I forms are situated at $\Omega_1 \ne 0$.  (b) Planar projections of four
helical polymorphic forms of a filament with a fixed contour length: normal
(blue),  coiled (brown), semicoiled (green), and curly-I (red).  (c) A
schematic of the model bacterium used in our simulations.  The cell body and
part of the discrete flagellum with a discretization length $h$ (drawn out of
proportion) are shown.  Other components are as described in the text.  (d) Two
snapshots from our simulations with a CW rotated flagellum: (top) a growing
semicoiled domain (green) evades the normal state (blue) and (bottom) a growing
curly-I domain (red) evades the semicoiled state (green).}

\label{schematic_model}
\end{center}
\end{figure}

We treat the flagellum as a slender body and parametrize its centerline
$\rb(s)$ by the contour length $s$ \cite{reinhardEPJE2010}.  The conformation
of a distorted flagellum including twist deformations is characterized by
orthonormal material tripods $\left\{\eb_1(s), \eb_2(s), \eb_3(s)\right\}$ at
each point on the centerline, where $\eb_3$ is the local tangent to the
centerline at $s$ and unit vectors $\eb_1$ and $\eb_2$ point along the
principal axes of the flagellar cross section.  The rotational strain vector
$\Omegab$ in
\begin{eqnarray}
\partial_s \eb_{\nu} = \Omegab \times \eb_{\nu}, \label{eqn_Omega}
\end{eqnarray}
with $\nu=1,2,3$, transports the material tripod along the centerline.
Therefore, together with the position $\rb(s=0)$ of one flagellar end,
$\Omegab(s,t)$ carries the complete information for the conformation of the
flagellum at time $t$.  Conversely, given the conformation in terms of the
centerline $\rb(s)$ and the material tripods at each $s$, the rotational strain
vector $\Omegab$ can be worked out from Eq.  (\ref{eqn_Omega}).  Introducing
the angle $\alpha$ between $\eb_1$ and the local normal to the centerline,
${\bf n} \propto \partial_s^2 \rb(s) $, one can relate the components
$\Omega_1, \Omega_2$ and $\Omega_3$ of $\Omegab$ with respect to the local
material frame to the curvature $\kappa$ and torsion $\tau$ of the flagellum
\cite{reinhardEPJE2010}:
\begin{eqnarray}
\Omega_1 = \kappa \sin{\alpha}, \quad\enspace 
\Omega_2 = \kappa \cos{\alpha}, \quad\enspace 
\Omega_3 = \tau + \partial_s \alpha. 
\end{eqnarray}

In 
harmonic approximation, the energy required per unit length of the
flagellum to induce small deformations $d\Omegab = \Omegab - \Omegab^{(n)}$
from the stable ground state $\Omegab^{(n)}$, is the Kirchhoff elastic free
energy density \cite{landau_elasticity}, 
\begin{eqnarray}
f_{\mathrm{K}} (\Omegab, \Omegab^{(n)}) 
          =   {A \over 2} \left[ (d\Omega_1)^2 + (d\Omega_2)^2 \right] 
            + {C \over 2} (d\Omega_3)^2, \label{eqn_f_K}
\end{eqnarray}
where $A$ and $C$ are the respective bending and twist rigidities of the
flagellum, assuming a circular cross section.  Assigning $f_{\mathrm{K}}
(\Omegab, \Omegab^{(n)})$ and a ground state energy $\delta_n$ to each
polymorphic form $\Omegab^{(n)}$, an extended version of the Kirchhoff free
energy density, capable of describing the multistability of a flagellum, is
formulated as \cite{reinhardPRL2013}
\begin{eqnarray}
f_{\mathrm{EK}} = \min_{\forall n} 
                  \left[f_{\mathrm{K}} (\Omegab, \Omegab^{(n)}) + \delta_n\right] 
                  + {A\over 2 } \xi^2 \left( \partial_s \Omegab \right)^2.
\label{eqn_f_EK}
\end{eqnarray}
Here, $\Omegab^{(n)}, n=1 \hyphen 4$, correspond, respectively, to the normal,
coiled, semicoiled, and curly-I forms of an \emph{E.\ coli} flagellum
\cite{reinhardPRL2013}.  We include the experimentally unobserved left-handed
coiled form that is intermediate in $\kappa$ and $\tau$ between the normal and
semicoiled forms and thus also in $\Omega$-space \cite{calladine1976}.  The
first term in Eq.  (\ref{eqn_f_EK}) implies that for any rotational strain
$\Omegab(s)$ at a given point $s$ on the flagellum, the elastic free energy
density is chosen to be that of the polymorphic form with the lowest energy.
The second term in Eq.  (\ref{eqn_f_EK}) enforces a smooth transition region of
width $\xi$ between two polymorphic domains.  

For an \emph{E. coli} flagellum, we use $A = 5.5\, \mathrm{pN}\mu\mathrm{m}^2$
and $C = 3.5\, \mathrm{pN}\mu\mathrm{m}^2$ \cite{turner2007}.  The stable
ground states $\Omegab^{(n)} \equiv \{\Omega_1^{(n)},\allowbreak
\Omega_2^{(n)},\allowbreak \Omega_3^{(n)}\}$ for the normal, coiled,
semicoiled, and curly-I forms are, respectively (in $\mu\mathrm{m}^{-1}$): $\{
0.00,\allowbreak 1.30,\allowbreak -2.11\}$, $\{-0.51,\allowbreak
1.74,\allowbreak -0.56\}$, $\{-1.18,\allowbreak 1.84,\allowbreak  0.98\}$, and
$\{-1.80,\allowbreak  1.56,\allowbreak 2.53\}$ \cite{reinhardPRL2013}.

The ground state energies $\delta_n$ in Eq. (\ref{eqn_f_EK}) are not known yet.
Their relative values determine both the height and shape of transition
barriers between consecutive minima [Fig.  \ref{schematic_model}(a)].  The
positions $\Omegab^{(n)}$ of the minima of $f_{\mathrm{EK}}$ are already fixed,
as are the parabolic shapes in the neighborhood of those minima.  Hence, the
full landscape of the extended Kirchhoff energy density is known once the
ground state energies $\delta_n$ or equivalently the transition barriers are
specified.

\section{\mbox{Equations of motion and numerical methods}}
\label{model}

\subsection{Dynamics of the flagella}

\subsubsection{Equations of motion}

We consider the dynamics of an elongated cell body with a flagellum emanating
from an arbitrary point on its surface and moving in an unbounded fluid of
viscosity $\eta$.  The equations of motion of the flagellum are given by the
Langevin equations for the dynamics of the centerline $\rb(s,t)$ and the twist
angle $\phi(s,t)$ about the centerline \cite{tapan_PRE2015}:
\begin{eqnarray}
\partial_t \rb &=& \mutens_t \left(\Fb_{\mathrm{el}} + \Fb_{\mathrm{s}} + 
                   \Fb_{\mathrm{th}}\right) + \vb_{\mathrm{h}} , \label{r_eqn} \\
\partial_t \phi &=& \mu_r (T_{\mathrm{el}} + T_{\mathrm{th}}). \label{phi_eqn}  
\end{eqnarray}
Here the $\Fb$'s and $T$'s are, respectively, the local forces and torques
acting on the flagellum due to elastic deformations, steric interactions, and
thermal noise. We will describe them below.  They are connected to the linear
($\partial_t \rb$) and angular ($\partial_t \phi$) velocities by the respective
self-mobilities   $\mutens_t = \eb_3 \otimes \eb_3/ \gamma_\parallel + (\Ib -
\eb_3 \otimes \eb_3)/ \gamma_\perp$ and $\mu_r = 1/\gamma_R$, where
$\gamma_\parallel = 1.6 \times 10^{-3}\, \mathrm{pNs}/\mu \mathrm{m}^2$,
$\gamma_\perp = 2.8 \times 10^{-3}\, \mathrm{pNs}/\mu \mathrm{m}^2$ and
$\gamma_R = 1.26 \times 10^{-6}\, \mathrm{pNs}$ are the anisotropic friction
coefficients per unit length for the flagellum of an \emph{E.\ coli}\,
\cite{reinhardEPJE2012}. Finally, $\vb_{\mathrm{h}}$ describes hydrodynamic
interactions between different parts of the flagellum as detailed below.

The elastic forces and torques follow from the total elastic free energy
$\mathcal{F} \left[\rb(s), \phi(s) \right]$ as:
\begin{equation} 
\Fb_{\mathrm{el}} = -\frac{\delta \mathcal{F}}{\delta \rb} 
\quad \mathrm{and} \quad
T_{\mathrm{el}} = -\frac{\delta \mathcal{F}}{\delta \phi}.
\label{phi_deriv} 
\end{equation} 
Here $\mathcal{F} \left[\rb(s), \phi(s) \right] = \int ds \left(f_{\mathrm{EK}}
+ f_{\mathrm{st}} \right)$, where $f_{\mathrm{EK}}$ is the Kirchhoff elastic
free energy density described in the previous section and $f_{\mathrm{st}} =
K(\partial_s \rb)^2/2$ with $K = 10^3 \mathrm{pN}$ is a stretching free energy
density introduced to prevent local stretching of the flagellum
\cite{reinhardPRL2013}.

\subsubsection{Discretization procedure, thermal noise, steric and hydrodynamic
interactions}

Before we address the other force and torque contributions in Eqs.\
(\ref{r_eqn}) and (\ref{phi_eqn}), we discuss the numerical scheme to update
the flagellar configuration in time.  In order to discretize the Langevin
equations, we consider discrete positions $\rb_i\equiv\rb(s_i)$ along the
flagellum and assign $\left\{\eb_1(i),\allowbreak \eb_2(i),\allowbreak
\eb_3(i)\right\}$ to the straight segment of length $h$ between $\rb_{i-1}$ and
$\rb_{i}$ [see Fig.\ \ref{schematic_model} (c)].  The forces now act on the
discrete points, while the torques are applied to the straight segments.  To
find the discretized versions of   $\Fb_{\mathrm{el}}$ and $T_{\mathrm{el}}$,
we discretize the derivatives in Eqs.\ (\ref{phi_deriv}) and write $\mathcal{F}
= \int ds \left(f_{\mathrm{k}} + f_{\mathrm{st}} \right)$ as a sum over the
segments. 

The thermal forces $\Fb_{\mathrm{th}}$ and torques $T_{\mathrm{th}}$ in Eqs.\
(\ref{r_eqn}) and (\ref{phi_eqn}), respectively, are negligible for the forward
propulsion of bacteria \cite{grahamPRE2011} but play an important role in the
polymorphic transformations during reversal of the flagellum
\cite{reinhardEPJE2012}. We, therefore, include them here.  As usual, they are
Gaussian random numbers with zero mean and variances $\langle
(F_{\mathrm{th}}^{\parallel/\perp})^2 \rangle = 2 k_B T
\gamma_{\parallel/\perp}/\Delta t$, where $\Delta t$ is the discrete time step
used in the simulation, and $\parallel$ and $\perp$ denote the directions along
and normal to the local tangent of the flagellum, respectively.  Similarly,
$\langle T_{\mathrm{th}}^2 \rangle = 2 k_B T \gamma_R/\Delta t$.

The steric force $\Fb_s$ enforces excluded-volume interactions among different
parts of the flagellum. This is needed in case distant flagellar parts try to
go through each other, e.g., during a strongly buckled state.  We model
the steric forces following Ref.\ \citenum{tapan_PRE2015}.


In order to model hydrodynamic interactions between different flagellar parts,
we treat each discrete point $\rb_i$ as a sphere of diameter equal to the
thickness of the flagellum.  Thus, we write $\vb_{\mathrm{h}} (\rb_i) =
\sum_{j\ne i} \mutens_{ij} \Fb(\rb_j)$, where the summation runs over all
points of the flagellum. Here $\mutens_{ij}$ is the Rotne-Prager mobility
matrix \cite{dhont} for spheres at $\rb_i$ and $\rb_j$ and $\Fb(\rb_j)$ is the
force acting at $\rb_j$. To be consistent with the picture that the spheres are
parts of the discretized flagellum, we neglect the hydrodynamic influence of
their rotation \cite{reichert_thesis}. Furthermore, to avoid huge computational
expenses, we also neglect any correlations between thermal forces
$\Fb_{\mathrm{th}}$ acting on different points, which occur due to hydrodynamic
interactions.

\subsection{The cell body, rotary motor, and flagellar hook}

The elongated cell body is modeled as a spherocylinder of length
$L_b=2.5\,\mu\mathrm{m}$ and width $d_b = 0.8\,\mu\mathrm{m}$ \cite{berg_ecoli,
tapan_PRE2015} [see Fig.\ \ref{schematic_model}(c)]. A flagellum with total
contour length $L$ is attached to
the point $\rb_{0}$ on the surface of the cell body. To represent a typical
flagellum undergoing a reverse rotation, we choose $\rb_{0}$ to be on the
cylindrical surface.  Flagella of an \emph{E. coli} are distributed randomly
over the entire cell body \cite{berg_ecoli}. This implies that an arbitrarily
chosen flagellum is more likely to be found on the cylindrical surface that has
a larger area compared to that of the spherical ends of the cell body.

A motor tripod $\left\{\eb_1(0), \eb_2(0), \eb_3(0) \right\}$ is introduced at
$\rb_{0}$, where $\eb_3(0)$ coincides with the shaft of the rotary motor
driving the flagellum.  A motor torque $\Tb_m = T_m \eb_3(0)$ drives the
flagellum by rotating this tripod.  The main part of the flagellum is coupled
to the motor tripod through the Kirchhoff elastic free energy density
$f_{\mathrm{k}}$ with a bending rigidity $A \to A_h = 2.0\, \mathrm{pN}
\mu\mathrm{m}^2$, and a twist rigidity $C \to C_h = 0.1 \times 10^{-2}
\mathrm{pN} \mu\mathrm{m}^2$\, \cite{sen2004, tapan_PRE2015}.  Thus, the
flagellum is connected to the motor shaft through a `hook' that acts like a
universal joint with low bending and high twist rigidities
\cite{hook_universaljoint} and allows the first flagellar segment along
$\eb_3(1)$ to be at any angle to the motor shaft and yet efficiently
transferring the driving torque to the flagellum. 

The cell body translates and rotates with velocities given, respectively, by 
\begin{equation}
\vb_b   = \mutens_b^t \Fb_b \quad \mathrm{and} \quad
\omegab_b = \mutens_b^r (\Tb_b + \Tb_m),
\end{equation}
where $ \Fb_b$ and $\Tb_b$ are the net force and torque acting on the center of
mass of the cell body.  They result from the force $\Fb_{\mathrm{el}} +
\Fb_{\mathrm{s}}$ that acts on the flagellar anchoring point. For simplicity,
the mobilities $\mutens_b^t$ and $\mutens_b^r$ are assigned the analytically
available values for a prolate spheroid of aspect ratio $L_b/d_b$
\cite{kim_karilla}. As in Ref.\ \citenum{tapan_PRE2015} the angle between
$\eb_3(0)$ and the long axis of the cell body is set to $55\degree$ to tune the
ratio for the bundle-to-body rotation rates during forward propulsion, to the
experimentally observed range \cite{turner2007}.

Finally, the excluded volume interaction between the cell body and the
flagellum is again modeled as in Ref.\ \citenum{tapan_PRE2015}.

\section{Results}
\label{results}

\subsection{\mbox{Effect of barrier heights: transition from semicoiled state}}

As pointed out earlier, since the positions of the minima of our model elastic
free energy are already fixed by the polymorphic states of the flagellum, the
full energy landscape is determined once the transition barriers between
consecutive minima are specified.  Within our model, the barriers are
determined by the differences in the ground-state energies $\{\delta_1,
\delta_2, \delta_3, \delta_4\}$, which we vary in the following, in order to
test how the free energy landscape affects the flagellar dynamics.  Since only
the relative heights of the minima are important, we set $\delta_1 = 0$.

As a starting point for a systematic study, we investigate a single transition
barrier connecting two consecutive minima.  In \emph{E.\ coli} a typical
transition between two neighboring polymorphic forms occurs between the
semicoiled and curly-I state.  So, we take $\delta_2 =\delta_3 = 0$  and vary
$\delta_4$.  We start with the flagellum entirely in the semicoiled form
(minimum 3) and apply a motor torque in the CW sense (as viewed from outside
the cell) for a duration comparable to the average tumbling time of $\sim 1\,
\mathrm{s}$.  The magnitude of the torque is fixed to a constant value of
$T_m =-3.0\, \mathrm{pN}\mu\mathrm{m}$, in agreement with experimental values
\cite{berg_ecoli}.  A typical snapshot from our simulations for this case is
shown in Fig. \ref{schematic_model}(d) (bottom).

\begin{figure}
\includegraphics[width=1.0\columnwidth]{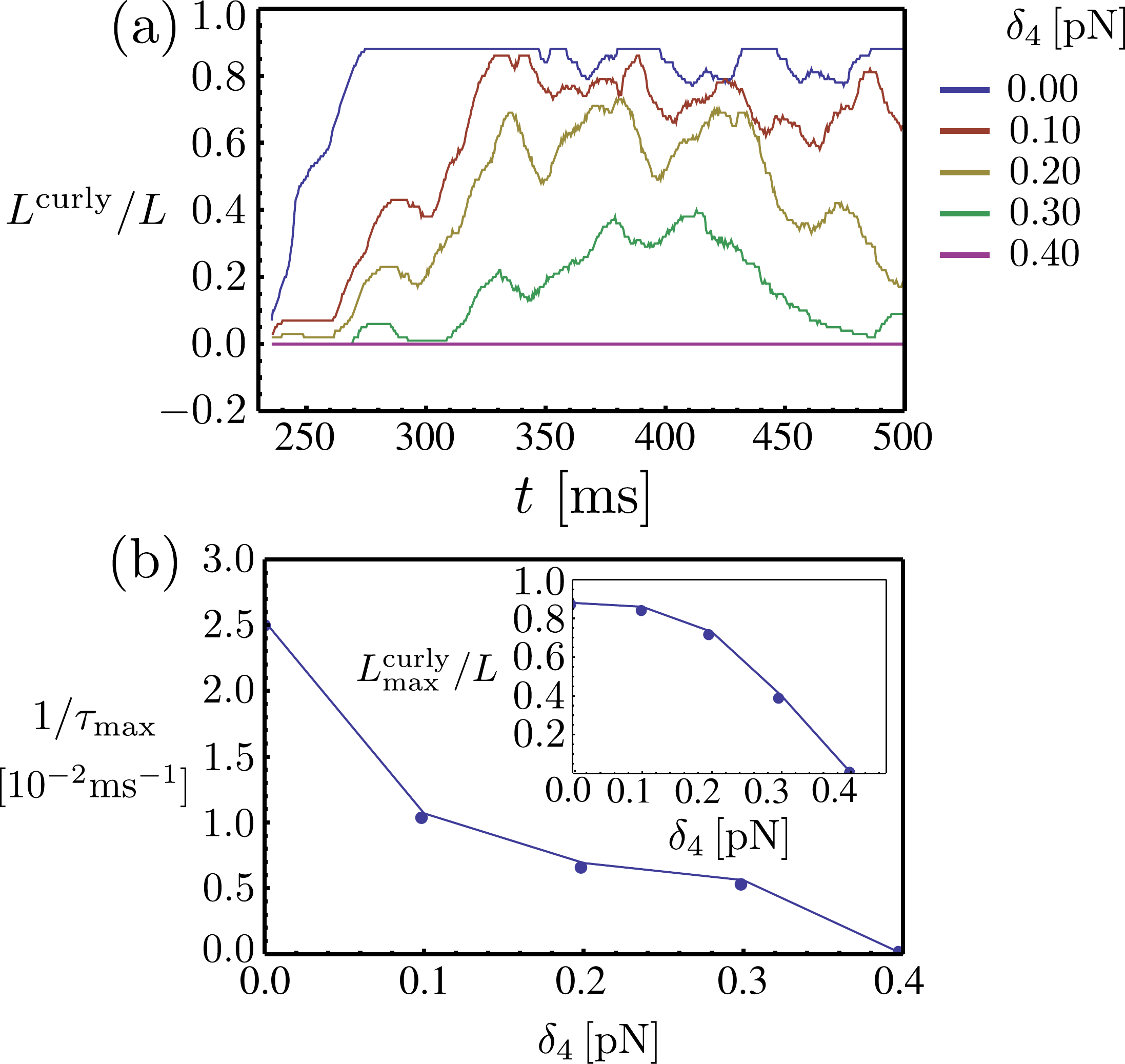}

\caption{Transition from semicoiled to curly-I form studied for different
ground-state energies $\delta_4$ (in $\mathrm{pN}$) and $\delta_1 =\delta_2 =
\delta_3 = 0.0\, \mathrm{pN}$.  (a) Time evolution of the fraction of the
flagellum in the curly-I state, $L^{\mathrm{curly}} /L$.  (b) Transition rate
$1/\tau_{\mathrm{max}}$ plotted versus $\delta_4$, where $\tau_{\mathrm{max}}$
is the time to achieve maximum length $L^{\mathrm{curly}}_{\mathrm{max}}$ in
the curly-I state.  Inset: $L^{\mathrm{curly}}_{\mathrm{max}}/L$ as a function
of $\delta_4$.  
}

\label{semi_to_curly}
\end{figure}

Our observations on the nature of the semicoiled-to-curly-I transition and the
stability of the curly-I form are summarized in Fig. \ref{semi_to_curly}.  In
graph (a) we plot the fraction of the flagellum in the curly-I state,
$L^{\mathrm{curly}}/L$, versus time $t$.  Below $\delta_4 = 0.4\, \mathrm{pN}$
the flagellum always transforms to the curly-I form, since a nonzero
$L^{\mathrm{curly}}$ eventually appears as time progresses.  Stability of the
curly-I state increases with decreasing $\delta_4$ and, ultimately, for
$\delta_4 < 0.1\, \mathrm{pN}$,  after an initial rapid build-up,
$L^{\mathrm{curly}}$ remains close to its maximum value as long as the
flagellum is reversely rotated.  However, for $\delta_4 > 0.1\, \mathrm{pN}$
(and $<0.4\, \mathrm{pN}$) fluctuations in $L^{\mathrm{curly}}$ are huge.

These fluctuations are due to flagellar portions in the curly-I form
transforming back to the semicoiled state. As $\delta_4$ increases, the barrier
height for the return jump from curly-I to semicoiled decreases and local
elastic stresses built up on the curly-I portion are sufficient to induce the
return transition.  We find the localized elastic stresses to be due to the
enhanced buckling of the curly-I form. The clockwise rotated curly-I state with
its right-handed helical structure generates a thrust force towards the
tumbling cell body, which is hardly translating. Thus the highly flexible
curly-I form buckles more easily and produces localized elastic stresses.

Varying $\delta_4$ also affects the maximum length
$L^{\mathrm{curly}}_{\mathrm{max}}$ of curly-I form, attained along the
flagellum during each run, and the transition time $\tau_{\mathrm{max}}$ to
reach $L^{\mathrm{curly}}_{\mathrm{max}}$. A continuous increase in the length
$L^{\mathrm{curly}}_{\mathrm{max}}$ is observed, when $\delta_4$ decreases
below the threshold value $0.4\, \mathrm{pN}$ [inset, Fig. 2(b)].  This
resembles the behavior of an order parameter characterizing a continuous phase
transition.  Furthermore, the transition time $\tau_{\mathrm{max}}$ increases
or the transition rate $1/\tau_{\mathrm{max}}$ decreases with growing
$\delta_4$, indicating a slowed-down transition to the curly-I form with
maximum length [Fig. \ref{semi_to_curly}(b)]. 

\subsection{Effect of barrier heights: transition from normal state}

\begin{figure}
\includegraphics[width=1.0\columnwidth]{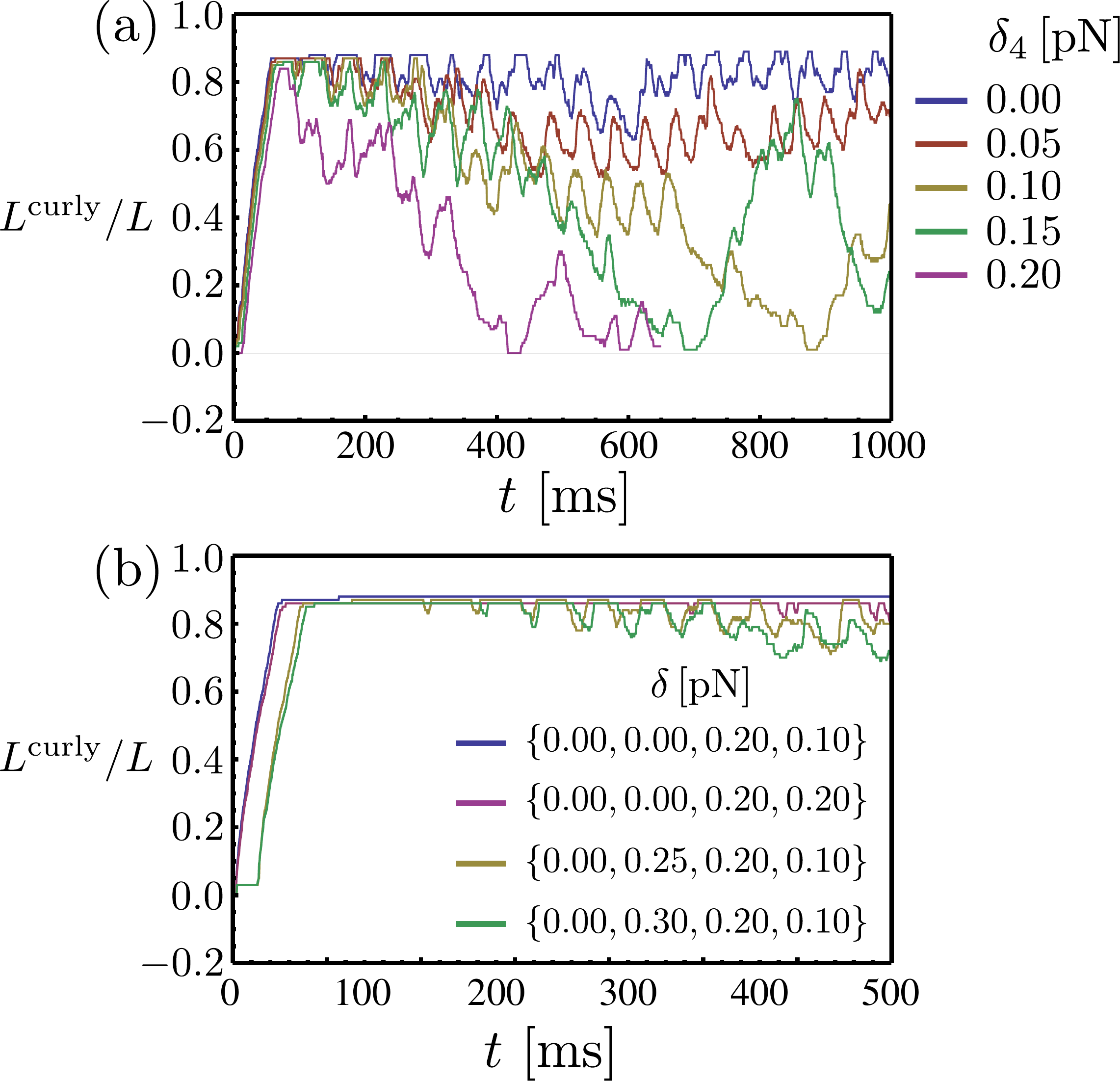}

\caption{ Transition from the normal form under reverse rotation.  (a) Time
evolution of the fraction of the flagellum in the curly-I state,
$L^{\mathrm{curly}} / L$, for various values of $\delta_4$ (in $\mathrm{pN}$)
and $\delta_1 =\delta_2 = \delta_3 = 0.0\, \mathrm{pN}$.  (b)
$L^{\mathrm{curly}} / L$ plotted against time for selected values of $\delta_2,
\delta_3$ and $\delta_4$.  } 

\label{normal_to_curly} 
\end{figure}

Having established the importance of changing the height of one transition
barrier, we now turn to the more complex problem, which is to test the dynamics
of the flagellum under reverse rotation starting from the normal state. Now all
three transition barriers, i.e., all three ground state energies $\delta_2,
\delta_3$, and $\delta_4$ become relevant. We vary them systematically as
explained below.  We start with a flagellum entirely in the normal form and
apply a motor torque, $T_m=-3.0\, \mathrm{pN}\mu\mathrm{m}$, in the CW sense for
a time similar to the average tumbling duration. In the following, we first
summarize the nature of the transition and the stability of different states
and then discuss distinct dynamic phases that emerge out of our observation.    

\subsubsection{Nature of transition and stability}

First we set $\delta_2 = \delta_3 = 0\, \mathrm{pN}$ and observe the time
evolution of $L^{\mathrm{curly}} / L$ as we vary $\delta_4$ [Fig.  3(a)].  We
find that  $L^{\mathrm{curly}}$ begins to increases from $t= 0\, \mathrm{ms}$
and reaches its maximum very early for each $\delta_4$. This implies that the
flagellum directly transforms into the curly-I state without residing in a
full-length semicoiled state as observed for real bacteria.  Moreover, the
final curly-I state is highly unstable against fluctuations for any non-zero
$\delta_4$. 

We find that these fluctuations in $L^{\mathrm{curly}}$ can be greatly reduced,
when $\delta_3$ is also shifted up [blue and violet curve, Fig.
\ref{normal_to_curly}(b)].  This becomes evident when comparing the
corresponding curves of Fig. 3(a) and Fig. 3(b) for the same values of
$\delta_4$ but with $\delta_3 = 0\, \mathrm{pN}$ and $\delta_3 = 0.2\,
\mathrm{pN}$, respectively. As explained in the previous section, an increase
in $\delta_4$ reduces the barrier height for leaving the curly-I state.  A
simultaneous increase in $\delta_3$, however, restores the barrier height and
thereby stabilizes the final curly-I form.

All curves in Fig. \ref{normal_to_curly}(a) and Fig.  \ref{normal_to_curly}(b)
(with $\delta_2=0$) attain approximately the same maximum value at nearly the
same time. Moreover, for $\delta_3$ or $\delta_4$ $> 0.35\, \mathrm{pN}$ a
transition to the curly-I state does not occur at all [see Fig.\
\ref{state_diagram}(b)].  This implies that $L^{\mathrm{curly}}_{\mathrm{max}}$
jumps from zero to a non-zero value at the threshold $\delta_3 \approx \delta_4
> 0.35\, \mathrm{pN}$.  Similarly, $\tau_{\mathrm{max}}$ also
remains nearly constant during the transition from the normal to the curly-I
state.  This behavior is unlike the transition from the semicoiled to the
curly-I state, where both $L^{\mathrm{curly}}_{\mathrm{max}}$ and
$\tau_{\mathrm{max}}^{-1}$ continuously decrease to zero when increasing
$\delta_4$ to $0.4\, \mathrm{pN}$ [see Fig.  \ref{semi_to_curly} (b)].  Note
while the semicoiled and curly-I states are separated by a single barrier, the
flagellum has to be pass three barriers during the transition from the normal
to the curly-I state. This might explain the discontinuous transition.

Next, we examine the effect of non-zero values of $\delta_2$. A comparison of
the time evolution of $L^{\mathrm{curly}} / L$ for the same set of $\delta_3,
\delta_4$ values but with different values of $\delta_2$ is shown in Fig. 3(b)
(yellow and green curves). It is clear that a non-zero but small value of
$\delta_2$ does not affect the above results qualitatively.  The nature of the
transition from the normal form and the stability of the curly-I state remain
the same as before as long as $\delta_2$ is small.  

This shows that in our quest of identifying the influence of the free energy
landscape on the polymorphic transformation, the most important parameters to
study are $\delta_3$ and $\delta_4$.  In the following we therefore ignore any
variation in $\delta_2$.

\subsubsection{Dynamic states exhibited by a flagellum under reverse rotation}

According to our findings, we take $\delta_2 = \delta_1 = 0$ and explore the
dynamic behavior of the flagellum as we vary $\delta_3$ and $\delta_4$. After a
thorough examination, six distinct dynamic states emerged, which are listed in
Fig. 4 in panels I-VI, respectively.  Each panel plots for specific values of
$\delta_3$ and $\delta_4$ the color-coded polymorphic forms spread along the
flagellum as the flagellum evolves in time $t$. The
distinct dynamic states are characterized as follows. 

(I) Direct transition from normal to curly-I state, which nearly spreads along
the whole flagellum and is dynamically very stable. 

(II) Transition to a dynamically stable curly-I state, which is accompanied by
small transient regions of semicoiled form, which appear and disappear locally. 

(III) Transition to a dynamically stable curly-I state with large transient
regions of semicoiled form.  The semicoiled domains appear 
repeatedly over time near the cell body ($s=0$) and move towards
the free end of the flagellum, where they shed off.

(IV) An initial nearly full-length transition to the curly-I state, followed by
a reverse transition to a full-length, relatively stable semicoiled state. 

(V) Transition to a dynamically stable semicoiled form, the curly-I state is
not reached.

(VI) Dynamically stable normal form. No significant flagellar portion
transforms to other polymorphic forms.

\subsubsection{Energy landscape and state diagram}
\label{para:state_diag}

\begin{figure}[h]
\includegraphics[width=1.0\columnwidth]{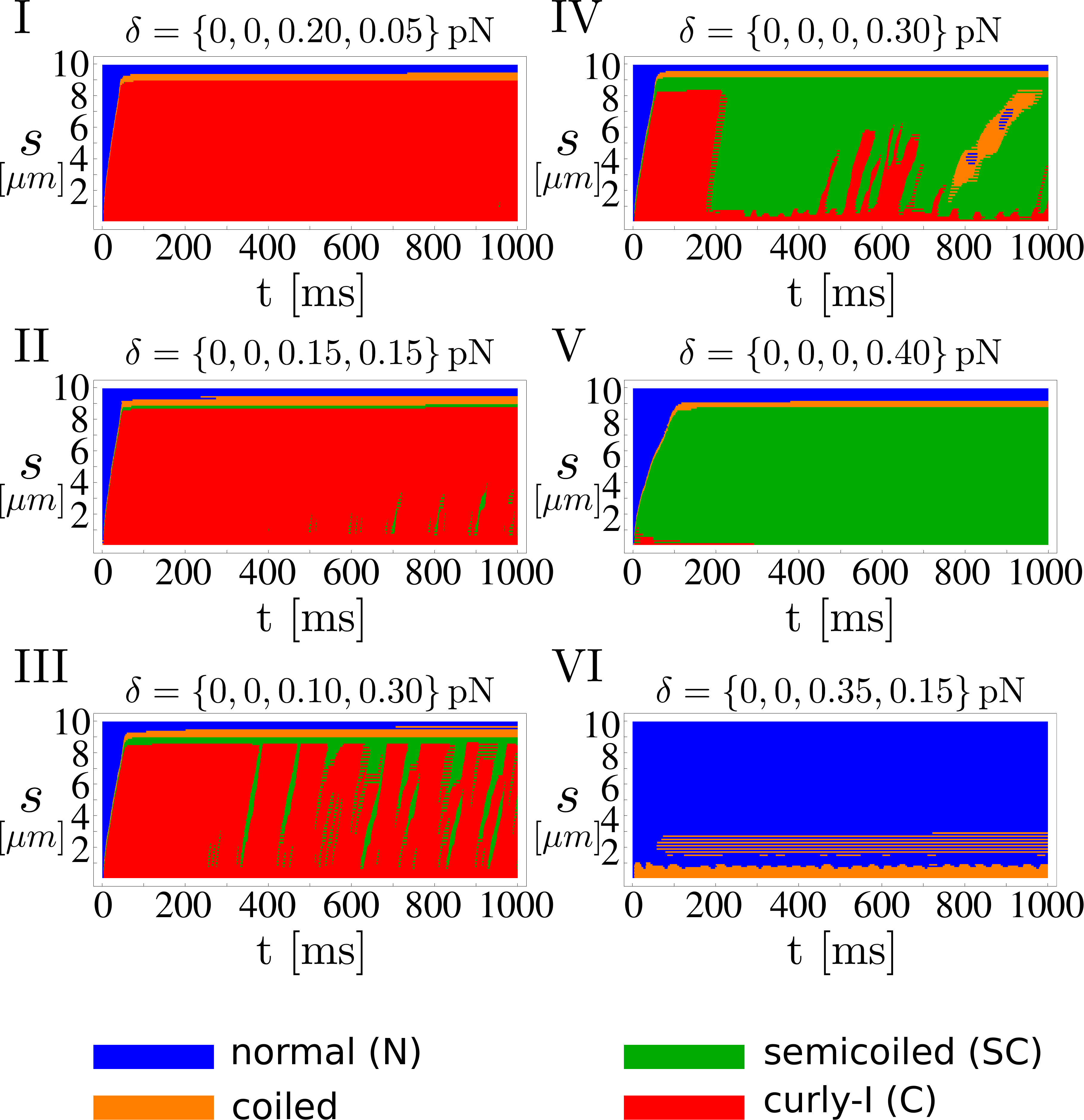}

\caption{Distinct dynamical states exhibited by a flagellum under reverse
rotation starting from the normal state. Panels I-VI show how the polymorphic
forms indicated by different colors evolve in time along the fla\-gellum.
Arclength $s$ gives the position on the flagellum.  For $\delta_1 = \delta_2 =
0$, typically observed dynamic states are: (I) dynamically stable C state, (II)
stable C state with transient SC regions, (III) repeated emergence of large
transient SC regions, (IV) reverse transition from curly-I to a full-length,
relatively stable SC state, (V) stable SC state, and (VI) stable N state.  See
text for a detailed description.  }

\label{dynamic_phases}
\end{figure}

\begin{figure}[h]
\includegraphics[width=0.95\columnwidth]{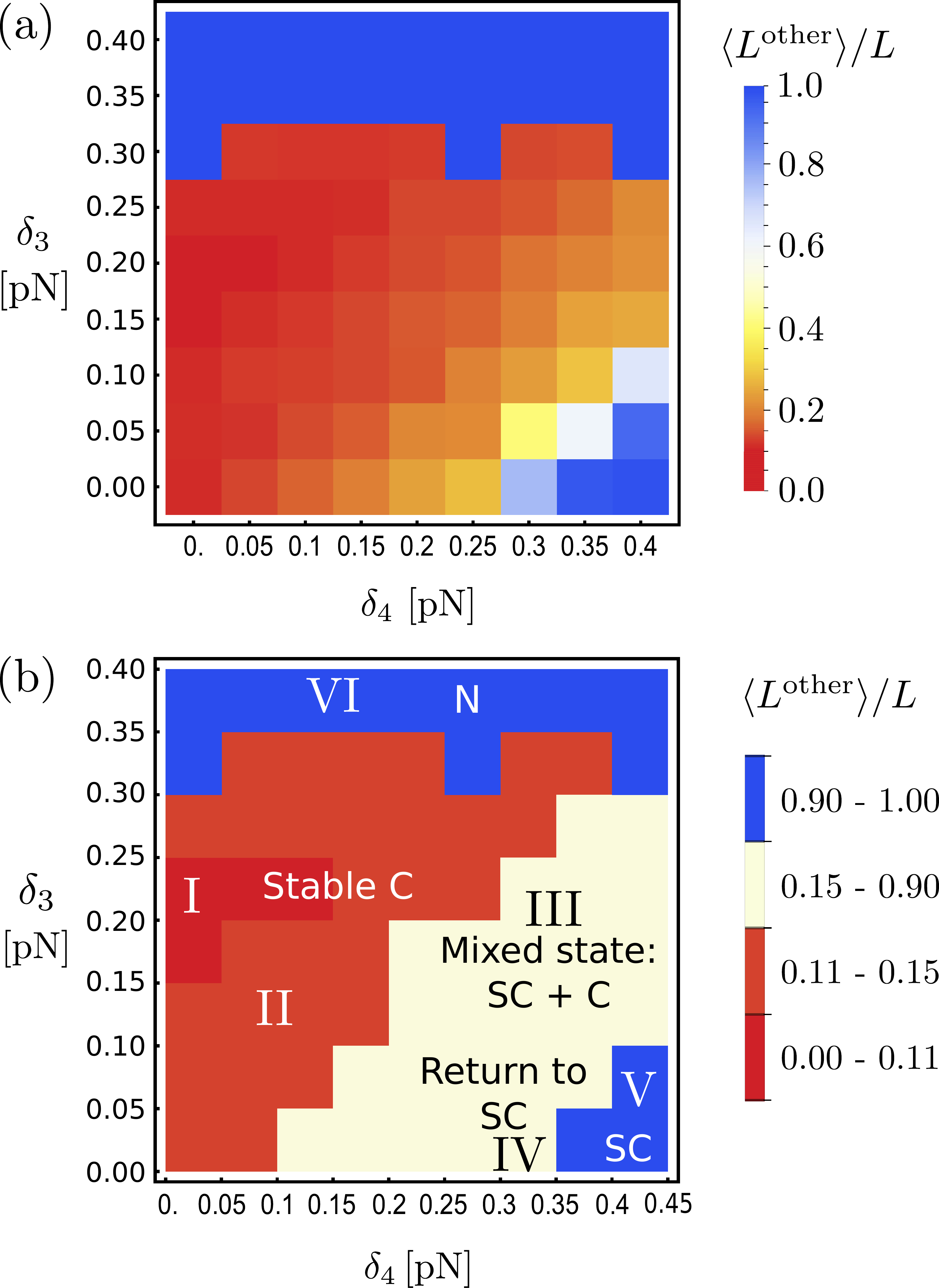}

\caption{Dynamic stability of curly-I state and state diagram for $\delta_1 =
\delta_2 = 0\, \mathrm{pN}$.  (a) Mean length of the flagellar portion,
which is not in the curly-I form, $\langle L^{\mathrm{other}} \rangle /L$,
represented in the $\delta_3$-$\delta_4$ plane by a color code.  (b) Colored
regions in the $\delta_3$-$\delta_4$ plane represent different length ranges of
$\langle L^{\mathrm{other}} \rangle$ in units of the total flagellar length;
dark red: $0$-$11\%$, light red: $11$-$15\%$, beige: $15$-$90\%$, and blue: $>
90\%$.  Roman numbers indicate the distinct dynamic phases shown in Fig. 4.}

\label{state_diagram} 
\end{figure}

Combining information obtained from a range of values for $\delta_3$ and
$\delta_4$, we finally arrive at a comprehensive picture of the flagellar
dynamics under reversal.  Figure 5(a) shows how the  average length $\langle
L^{\mathrm{other}} \rangle$ of the flagellar portion, which is not in the
curly-I state, varies in the $\delta_3 \hyphen \delta_4$ plane. Here $\langle
\cdots \rangle$ represents an average over time excluding the initial period of
transition from the normal form. So, a low value of $\langle L^{\mathrm{other}}
\rangle$ implies a dynamically stable curly-I state, whereas a high value
implies either large fluctuations in the curl-I state or no transition to this
state.  We already recognize that the curly-I state is unstable for high values
of both $\delta_3$ and $\delta_4$.  However, more interestingly, we also notice
that a very stable curly-I state occurs even for higher values of $\delta_4$
provided we increase $\delta_3$  accordingly.

Going into depth, we now monitor the detailed time evolution of the polymorphic
states along the whole flagellum (like in Fig. 4) for the complete $\delta_3
\hyphen \delta_4$ plane.  We identify all the dynamic states listed in the
previous section and find that states I-VI are characterized by values of
$\langle L^{\mathrm{other}} \rangle$ in the ranges $0 \hyphen 11\%$, $11
\hyphen 15\%$, $15 \hyphen 90\%$ and $> 90\%$  of the flagellar length.  The
respective regions in the $\delta_3 \hyphen \delta_4$ plane are extracted from
Fig. \ref{state_diagram}(a) and represented by colors in Fig.
\ref{state_diagram}(b).  The corresponding dynamic states are indicated by
their Roman numbers.  Note that states III and IV are not distinguishable from
each other by the range defined for $\langle L^{\mathrm{other}} \rangle$. The
same holds for states V and VI.  However, the full time evolution of the
polymorphic forms along the flagellum clearly identifies them as separate
states and we mark their occurrence in the $\delta_3 \hyphen \delta_4$ plane.

\section{Discussion}
\label{disscn}

A comparison of our findings with experimental results should, in principle,
give the unknown parameters of our extended Kirchhoff free energy.  It should
also reveal how accurately the elastic properties of the flagellum are
described by this energy.

The relevance of our findings becomes clear, when we consider the full time
evolution of a flagellum under reversal of the driving torque as observed in
experiments \cite{turner2000, turner2007}.  For a real \emph{E. coli}, a
reversely rotated flagellum first transforms in full length to the semicoiled
form, followed by the curly-I form, which persists until the rotation switches
back to the CCW sense.  Even though the appearance of the semicoiled domain is
brief, it extends over the full flagellum and does not fluctuate into other
forms until the curly-I domain grows from the cell body and takes over the
whole flagellum.  So, to model the flagellar dynamics under reversal correctly,
the intermediate full-length semicoiled state should occur.

However, we do not observe such a behavior in the dynamic states reported in
Fig. 4. The flagellum either transforms directly into the curly-I form [regions
I and II in Fig. \ref{state_diagram}(b)] or it remains in the semicoiled state
[region V in Fig.  \ref{state_diagram}(b)].  Thus, for any combination of
$\delta_3$ and $\delta_4$ there is never an intermediate transition to the
full-length semicoiled state followed by an automatic transition to the curly-I
form. This result of our model clearly is in contrast to what is often observed
for real bacteria. Non-zero but small values of $\delta_2$ should not change
this behavior because, as we have shown earlier, $\delta_2$ does not affect the
dynamics in any significant way.

One possible reason for this discrepancy with experiments is the harmonic
approximation of the Kirchhoff free energy in the deformation $d\Omegab$.
For large deformations anharmonic terms become important. Inclusion of
higher powers of $d\Omegab$ in the Kirchhoff free energy would result in a
highly non-trivial energy landscape, which is expected to modify the flagellar
dynamics observed in our present study. Possible reasons for inclusion of extra
terms might also be related to the finer details of the molecular structure of
the flagellum and the hook, not investigated fully so far. Effects of such
extra terms in the Kirchhoff free energy, allowed by symmetries, will be
examined in the future.

An alternative way to capture the experimental pathway of the observed
polymorphic transitions in \emph{E.\ coli} within the present model is the
following.  There could be an internal switch, realized by some biological
mechanism, that causes an effective jump in the $\delta_3 \hyphen \delta_4$
plane during the dynamics. Thus, the free energy in the first period of the
flagellar reversal corresponds to the region V of Fig.\ 5(b), until the
transformation to the semicoiled form  is complete.  Then, the values of
$\delta_3, \delta_4$ switch to the region I+II for the remaining part of the
dynamics as long as the reversal continues. 

\section{Conclusions}
\label{concl}

In conclusion, we examined the dynamics of a reversely rotated \emph{E. coli}
flagellum attached to a moving cell body by thoroughly exploring the elastic
free energy landscape of the flagellum.  We considered a general form of the
extended Kirchhoff free energy that was shown to be most appropriate for both
stretching and rotation-induced polymorphism. Minima of this free energy
correspond to the known polymorphic forms of the flagellum.  However, the
relative values of the ground state energies of those minima are not known.

We systematically studied how changes in the ground state energies influence
the transition of a flagellum in two cases: from a semicoiled to the curly-I
state, and from a normal to the curly-I state, respectively. These transitions
are relevant during flagellar reversal.

We find that under reverse rotation, a normal flagellum can transform to a
curly-I state, whose stability depends sensitively on the relative ground state
energies of the involved polymorphic forms.  The transition to the curly-I form
can even be forbidden, making the flagellum either to continue in the normal
form or to transform to a stable semicoiled state.  We have classified these
distinct dynamical states and obtained a state diagram for varying ground state
energies.  From this, one infers that for any combination of the ground state
energies in our model, an intermediate transition to a full-length semicoiled
state followed by a transition to the final curly-I form cannot be realized.
However, we suggest an alternative way to reproduce within our model such a
sequence of transitions observed for a real bacterium.

Our study provides a complete picture of how the elastic free energy landscape
determines the dynamics of a reversely rotated flagellum attached to a movable
cell body.  Therefore, our findings are important for the proper modeling of
the locomotion of a bacterium including its tumbling.  We show that the full
phenomenology of an \emph{E. coli} flagellum cannot be realized by simply
adjusting the parameters of the extended Kirchhoff free energy.  This calls for
alternative approaches. On the one hand, investigating the importance of
anharmonic terms in the free energy or how finer details of the hook
\cite{stocker_hook2013} influence the flageller dynamics could be two
possibilities in this direction.

On the other hand, based on the established state diagram we suggested an ad
hoc method to realize the correct polymorphic sequence of an \emph{E.\ coli}
flagellum.  Implementing this method allows a thorough theoretical
investigation of the complex and still not fully understood tumbling event of
an \emph{E.  coli}.  Moreover, since the elastic properties of bacterial
flagella are similar, our method can also be applied to explore the sequence of
polymorphic forms seen in other peritrichous bacteria during tumbling.

\section*{Acknowledgement}

We thank the VW foundation 
for financial support within the program "Computational Soft Matter and
Biophysics" (Grant No. 86 801)







\providecommand*{\mcitethebibliography}{\thebibliography}
\csname @ifundefined\endcsname{endmcitethebibliography}
{\let\endmcitethebibliography\endthebibliography}{}

\end{document}